# Dense and single-phase $KTaO_3$ ceramics obtained by spark plasma sintering

L. Féger[1,±], F. Giovannelli[1], G. Vats[2,3], J. Alves[1], B. Pignon[1], E. K. H. Salje[4], I. Monot-Laffez[1], G. F. Nataf[1,*]

± lucile.feger@univ-tours.fr

\* guillaume.nataf@univ-tours.fr

[1] GREMAN UMR7347, CNRS, University of Tours, INSA Centre Val de Loire, 37000 Tours, France

[2] Groningen Cognitive Systems and Materials Center (CogniGron), University of Groningen, Groningen 9747 AG, The Netherlands

[3] Department of Physics and Astronomy, Katholieke Universiteit Leuven, Celestijnenlaan 200D, 3001 Leuven, Belgium

[4] Department of Earth Sciences, University of Cambridge, Cambridge CB2 3EQ, UK

## Abstract

Potassium tantalate ($KTaO_3$) is a promising material for dielectric applications at low temperature. However, dense and single-phase ceramics cannot be obtained by conventional sintering because of the evaporation of potassium that leads to secondary phases. Here, we demonstrate that spark plasma sintering is a suitable method to obtain dense and single-phase $KTaO_3$ ceramics, by optimizing three parameters: initial composition, temperature, and pressure. A 2 mol% K-excess in the precursors leads to a large grain growth and dense single-phase ceramics. Without K-excess, a small amount of secondary phase ($K_6Ta_{10.8}O_{30}$) is observed at the surface but can be removed by polishing. At ~10 K, the dielectric permittivity is 4 times higher in the ceramic from the 2 mol% K-excess powder, because of the larger grain size. The thermal conductivity decreases with decreasing grain size and stays above the thermal conductivity of $KNbO_3$ ceramics.

## 1. Introduction

Since the development of new technologies in optoelectronic, high frequency telecommunication and holographic fields, tantalate perovskites have been at the centre of attention because of their flexible dielectric properties. This is the case of potassium tantalate ($KTaO_3$), an incipient ferroelectric, whose dielectric permittivity $\varepsilon'$ increases when temperature decreases, while keeping low dielectric losses. These properties make this perovskite suitable for low-temperature applications, e.g. for tunable microwave components [1,2]. $KTaO_3$ attracts also much attention for its ability to generate a two-dimensional electron gas at its surface [3–5], and for the existence of defect-induced piezoelectricity in its nominally centrosymmetric phase [6,7], opening possibilities for new electromechanical devices and high-temperature piezoelectrics [7,8]. In parallel, alkaline-based perovskites (e.g. $KNbO_3$ and $NaNbO_3$) have been investigated for their low thermal conductivities compared to other perovskites [9,10], attributed to cation deficiency [9], with possible applications in the field of thermal barrier coatings or thermoelectric devices [9].

$KTaO_3$ has been mostly studied in the form of single crystals [11,12] and powders [13–15], but ceramics are preferred for commercial use. Unfortunately, the solid-state synthesis of alkaline ceramics is difficult because of highly hygroscopic precursors and the evaporation of potassium during conventional sintering. This leads to the nucleation of secondary phases [16–20] even when sintering is performed at relatively low temperature (e.g. 1598 K) [20]. A way to overcome this issue is to add in the precursor 2 to 5 mol% potassium-excess to compensate for the loss [18,21]. However, this leads to ceramics with densities below 88% [18]. The addition of dopants, such as $Mn_3O_4$ or $Fe_2O_3$, can help the densification process through the formation of a liquid phase (at 1523 K for $Mn_3O_4$), but it significantly alters dielectric properties [22–24]. Hot isostatic pressing (HIP) has shown promising results, with dense and stoichiometric ceramics [17,20], but this technology is not easy to handle and requires long process times.

As an alternative route, we propose to use spark plasma sintering (SPS) [25], a technique known to be effective on perovskite systems similar to $KTaO_3$, such as $(K,Na)NbO_3$ and $(K,Na)(Nb,Ta)O_3$, to reduce sintering time and temperature, which leads to high relative densities [26–29]. Here, we investigate different spark plasma sintering parameters, starting from 0 mol% excess to 2 mol% excess in potassium in the precursors' compositions. We obtain dense and single-phase ceramics of $KTaO_3$, with sintering times of a minute or less, and compare their structures and morphologies. We take advantage of the high relative density of our ceramics with respect to previous works, and of the absence of secondary phases in the ceramic without K-excess to discuss the role of grain size on the dielectric constant and thermal conductivity values.

2. Methods

$KTaO_3$ was prepared by the solid-state method. Two commercial high-purity powders, $K_2CO_3$ (99.9%, ChemPur) and $Ta_2O_5$ (99.9%, ChemPur), were used as starting materials for the synthesis. $K_2CO_3$ was dried overnight at 300°C and the mixture was then homogenized by wet ball milling (with ethanol), using a tungsten carbide bowl and balls with a diameter of 10 mm, for 5 hours at 200 RPM. The mixture was dried at 353 K overnight. Once dry, the powders were calcined in air in a tubular furnace at 1123 K for 5 hours at a heating and cooling rate of 2.5 K min$^{-1}$.

$KTaO_3$ powders were densified by spark plasma sintering using a SPS632Lx, Dr.Sinter (Fuji Electronics) equipment. Sintering was performed at different temperatures and pressures to reach high densities without K-loss. A short dwell time (0.5-1 min) and a heating rate of 100 K min$^{-1}$ were applied to minimize grain growth. The uniaxial pressure was applied on the pellet at ambient temperature, held during the heating and dwell process, and then released during the cooling step. Two different pressures (60 and 100 MPa) were used to observe the influence of this parameter on the densification of the samples. A graphite mould with a diameter of 10 mm and punches were used with about 600 mg of $KTaO_3$ powder. A graphite foil was placed inside the mould to prevent chemical reactions between the sample and the mould, and to ensure a good electrical contact. The upper punch displacement is measured with a digital *z*-axis displacement gauge (minimum 0.001 mm). A low vacuum (below 20 Pa) was maintained in the chamber to prevent oxidation and degradation of the graphite materials.

The pellets were then post-annealed for 1 hour under oxygen flow at 1273 K in a tubular furnace to oxidize them to compensate oxygen losses due to reducing conditions in the SPS process, and to burn out residual carbon.

Phase purity of the different powders and ceramics was controlled by X-ray diffraction using a Cu-Kα radiation filtered by a Ni foil (D8 Advance Bruker X-ray diffractometer). X-ray diffraction patterns were collected with a scanning speed of 2.3° min$^{-1}$ with a step size of 0.02° over a 2θ angular range from 10° to 80°, at room temperature. Rietveld refinements were performed with the FullProf software. Particle sizes of the powders were evaluated by laser granulometry (Horiba LA950-V2), with the Fraunhofer method. Densities of ceramics were calculated from the sample weights and volumes obtained from geometrical dimensions and with the Archimedes method using a MS204TS/00 analytical balance (Mettler Toledo) in distilled water for comparison. Relative densities were calculated considering a theoretical density of 7.02 g cm$^{-3}$ [30]. Samples microstructures were observed by scanning electron microscopy (SEM FEG - Tescan) with a secondary electron detector (acceleration voltage of 5 kV) on fractured samples. Complementary SEM images were taken with a backscattered electron detector on polished surfaces. The surfaces were polished with a Buehler EcoMet 30 polishing machine with a 250 nm diamond paste (Presi), under a force of 5 N. A Buehler 40-7220 microCloth and samples were rotating in opposite directions at 200 rpm and 40 rpm, respectively. Grain size analysis was performed on fractured and polished surfaces with the ImageJ software and using the SEM scale as a reference (1) by measuring 400 grains and (2) by the linear intercept method with 10 lines of 6 µm on each sample.

Low-temperature dielectric measurements were performed on dense and single-phase ceramics from powders with 0 mol% K-excess and 2 mol% K-excess. Polished faces of the ceramics were covered with silver paint as electrodes. Measurements were performed from 20 Hz to 2 MHz on cooling and heating between 300 K and 10 K, in 5 K steps, with a Quantum Design PPMS and an Agilent (Keysight) E4980 LCR meter.

Thermal diffusivity was measured on dense and single-phase ceramics from powders with 0 mol% K-excess and 2 mol% K-excess, with the laser flash setup (Netzsch LFA 457). A thin graphite coating was used to maximize the absorption of the laser light. The experiment was performed in air from 323 K to 973 K, in 50 K steps. Each sample was measured three times at each temperature. The specific heat capacity ($C_p$) was measured on crushed ceramics by differential scanning calorimetry (Netzsch STA 449 F3 Jupiter) from 330 to 1000 K, with a heating rate of 20 K min$^{-1}$ in platinum crucibles and nitrogen atmosphere.

The thermal conductivity was then calculated using equation 1:

$$\kappa(T) = \rho(T) \cdot \alpha(T) \cdot C_p(T) \quad (1)$$

With $\kappa$(T) the thermal conductivity of the sample in W m$^{-1}$ K$^{-1}$, $\rho(T)$ the measured density of the ceramics in kg m$^{-3}$ (note that in our case we used the room temperature density for the calculation, given that its temperature variations are negligible compared to the temperature variations of the other parameters [31]), $\alpha(T)$ the diffusivity in m² s$^{-1}$ and $C_p(T)$ the specific heat capacity in J kg$^{-1}$ K$^{-1}$.

## 3. Results and discussion

### 3.1 KTaO$_3$ powder with 2 mol% K-excess

X-ray diffraction confirms that the raw powder with an initial excess of 2 mol% in potassium is single-phase KTaO$_3$ [30] (Fig. 1a). Table 1 shows the optimal parameters for the spark plasma sintering. At 1223 K and 60 MPa, the relative density is of 94%, increasing to 97% at 1243 K and 60 MPa. This is consistent with shrinkage curves shown in figure 2 indicating the beginning of the densification at 950 K for the composition in K-excess (red curve). X-ray diffraction shows that the obtained ceramics are single phase on both sides (Fig. 1b, c). The maximum temperature of sintering is restricted by the existence of a eutectic between KTaO$_3$ and K$_3$TaO$_4$ melting around 1373 K [32]. Given that spark plasma sintering temperatures are often lower than for conventional sintering, this sets an upper limit of 1243 K for sintering.

Annealing in an oxygen atmosphere at 1273 K for 1 hour was optimized by TGA analysis (Supplementary Note 1) and found to be the most suitable thermal treatment to reoxygenate the ceramics. After this step, the initially grey pellets turned white, which is consistent with a higher oxidation state. X-rays still do not show diffraction peaks belonging to the K$_6$Ta$_{10.8}$O$_{30}$ secondary phase [33], confirming that there was less than 2 mol% loss of potassium during the process and that the ceramics are still single phase (Fig. 1d).

### 3.2 KTaO$_3$ powder with 0 mol% K-excess

X-ray diffraction confirms that the powder without excess in potassium is single phase (Fig. 3a). Two main trends appear in table 1 to increase the relative density of the ceramics: an increase in sintering temperature and an increase in applied pressure. As such, ceramics sintered at 1273 K and 60 MPa have a relative density of 77%, lower than for ceramics sintered at 1373 K and 60 MPa that have a density of 79%. Sintering at 1373 K and 100 MPa leads to a relative density of 86%.

At higher sintering temperatures (1423 K) and 100 MPa, the relative density reaches 95% and X-ray diffraction reveals the appearance of K$_6$Ta$_{10.8}$O$_{30}$ [33] on one surface of the ceramics, due to a small loss in potassium (Fig. 3b). Note that it is visible only on a logarithmic scale, as shown in Fig. 3. Polishing was enough to remove this secondary phase that forms at the surface. This secondary phase is not visible on the other surface (Fig. 3c). This phenomenon is consistent with the unidirectional charge flows created during the spark plasma sintering process. The unidirectional electric pulses going through the sample during sintering lead to a displacement of mobile charge carriers (electrons, oxygen ions $O^{2-}$, potassium ions $K^+$) [34,35]. These displacements create here a composition gradient in the pellet. The $K_6Ta_{10.8}O_{30}$ phase is detected on the upper surface (face A), consistent with a $K^+$ flow going backward with respect to the electron flow [34,35]. As seen in figure 2, for this powder, higher sintering temperatures were required, with shrinkage beginning at 1090 K instead of 950 K for the K-excess composition. After spark plasma sintering, ceramics had a black/blueish colour with a colour gradient due to a reduction of $KTaO_3$ stronger on one face than the other. This gradient disappears after annealing under an oxygen atmosphere (1273 K, 1h). Since the annealing was performed after removal of the secondary phase by polishing, both faces exhibit single-phase $KTaO_3$ (Fig. 3d).

Sintering at 1473 K resulted in a particularly high relative density of the ceramics (~97%) but X-ray diffraction revealed that the decomposition of $KTaO_3$ in $K_6Ta_{10.8}O_{30}$ was stronger and could not be removed by a light polishing (Supplementary Note 2).

*3.3 Influence of the initial composition on grain sizes*

As shown in figures 4a and 4c, a study of the particle size distribution of the powders was performed. For both syntheses, particle distributions were very similar: the median particle size was around 600 nm and the first decile was around 350 nm for both powders; the ninth decile was slightly higher for the 0 mol% K-excess powder than for the 2 mol% K-excess powder (1.9 µm and 1.8 µm, respectively). These results were confirmed by SEM observations (figures 4b and 4d). SEM images also reveal that the powder without K-excess forms large agglomerates with a diameter of 1 to 2 µm (figure 4b) that can artificially lead to an increase in the size of the particles measured by laser granulometry (figure 4a). In the case of the synthesis with 2 mol% K-excess, particles are more dispersed and less agglomerated (figure 4d). They also show a better crystallization with well-defined cubic shapes characteristic of $KTaO_3$, consistent with the slightly smaller full width at half maximum of the (110) diffraction peak in this powder (0.11° for the powder with 2 mol% K-excess and 0.15° for the powder with 0 mol% K-excess).

SEM observations were then performed on the dense sintered ceramics once fractured and afterwards on polished surfaces (Figure 5). They confirm the high relative density of both ceramics but reveal very different grain sizes. In figures 5a-d, the ceramics from the powder without K-excess shows grains similar in size to the powder (fig. 4b). A largely more significant grain growth is observed in the 2 mol% K-excess ceramics (figures 5e-h), where grains expand up to 10 µm with a large dispersion in size. This grain growth is consistent with the observations by Tkach *et al.* [18], who reported comparable grain growth for K/Ta ratios over 1 by conventional sintering, due to the presence of the eutectic and a liquid phase formation favouring both grain growth and densification at 1373 K. One can thus expect that the eutectic liquid phase has formed in the K-excess composition sintered at 1243 K and 60 MPa. In addition, there is a possible eutectic between $KTaO_3$ and $K_2CO_3$ melting around 1073 K [32]. Since it melts at lower temperature it would favour grain growth even more, but it requires the compounds to absorb $CO_2$ from the air, for which we have no evidence since we did not observe diffraction peaks related to $K_2CO_3$, neither before nor after sintering.

*3.4 Dielectric measurements*

As shown in figures 6a and 6b, the permittivity at ~10 K is 4 times higher in $KTaO_3$ from the powder with 2 mol% K-excess than in $KTaO_3$ from the powder without K-excess. This difference has been observed before in a study of the influence of K-excess on dielectric properties of ceramics obtained by conventional sintering and attributed to either a decrease of the grain boundary contribution to the dielectric permittivity for K/Ta > 1 [18] or to the disappearance of the less polarizable secondary phase. In our study, even the ceramics from the powder without K-excess is single phase (within the sensitivity of X-ray diffraction), contrary to ref. 18. This is a strong indication that the difference in dielectric permittivity between the two measured samples results from the different grain sizes of the ceramics. The K-excess-powder-based ceramics have larger grains than the stoichiometric ones and thus a higher permittivity. The value of the permittivity at 10 K is around 4000 for the 2 mol% K-excess ceramics, which is similar to the single crystal value [11,36], confirming that ceramics obtained by spark plasma sintering are suitable for low-temperature applications.

The permittivity of the 2 mol% K-excess ceramics remains higher than the permittivity of the 0 mol% K-excess ceramics up to 250 K, where it is still about 3 times higher. Such a large difference has not been observed before, even though the permittivity of the sample with initially 5 mol% K-excess in ref. 18 exhibited a higher permittivity than the sample with initially 0 mol% K-excess. Note that the relative density of our ceramics is higher than in their work (95-97% by Archimedes compared to 87-90%). Furthermore, in our work, the ceramics with 2 mol% K-excess can still have an excess of potassium, since there is little loss of K during the SPS process, which could be the cause of the increase in permittivity. A small amount of impurities, acting as co-dopants, could also contribute to this increase, as previously demonstrated for weakly co-doped $KTaO_3$ with Li and Mn [37].

Near room temperature, the permittivity of both samples increases with increasing temperatures and decreasing frequencies, as previously observed in $KTaO_3$ single crystals [38], which is typical of the presence of charged crystal defects that contribute to a space-charge polarization [39]. This type of polarization is usually building up at interfaces such as grain boundaries, which explains why the increase in permittivity is stronger in the ceramics with 0 mol% K-excess where grains are smaller. This is consistent with the difference in dielectric losses between both samples, near room temperature (figures 6c,d).

In the case of $KTaO_3$ without K-excess, there is a peak in dielectric losses at 45 K at 1 kHz. This peak is noticeable in almost every dielectric losses spectra of $KTaO_3$ single crystals and ceramics and its origin is still unknown [11,12,17,20,40].

The activation energy ($E_a$) of the peak in dielectric losses, is calculated using the Arrhenius's law (equation 2) following the evolution of the relaxation frequency $f$:

$$f = f_0 \cdot \exp\left(\frac{-E_a}{k_B T}\right) \quad (2)$$

Where $k_B$ is the Boltzmann constant, $T$ the temperature and $f_0$ a constant.

It gives $E_a(R0) = 85$ meV (Supplementary Note 4), consistent with values reported in the literature ranging between 70 and 90 meV [11,40–42].

In the case of $KTaO_3$ with 2 mol% K-excess, more peaks in the dielectric losses are observed: 25 K (R0'), 85 K (R1), 140 K (R2) and 220 K (R3) at 1 kHz, with activation energies of 40 mev, 150 meV, 300 meV and 480 meV, respectively (Supplementary Note 4). Several peaks in this temperature range have been reported before in ceramics of $KTaO_3$ [17, 18, 20, 21]. This is because even a very small amount of defects (impurities, vacancies, lines defects, etc.) will easily lead to local dipole structures with different order–disorder energies and give rise to new loss peaks. It has been shown before that the intensity and position of these peaks vary with the amount of K-excess in the ceramics [18].

*3.5 Thermal conductivity*

Figure 7a shows the thermal diffusivity of ceramics with and without potassium excess as a function of temperature. In both cases, the thermal diffusivity is decreasing with increasing temperature. The thermal diffusivity of $KTaO_3$ with 0 mol% K-excess is slightly lower than the thermal diffusivity of $KTaO_3$ with 2 mol% K-excess. Figure 7b shows the specific heat capacity of $KTaO_3$ samples with 0 mol% K-excess and with 2 mol% K-excess, as a function of temperature. They both exhibit a similar behaviour, with a specific heat capacity ranging from ~400 J K$^{-1}$ kg$^{-1}$ at 300 K to ~500 J K$^{-1}$ kg$^{-1}$ at 1000 K, in agreement with the literature [43]. The difference between both curves is within the margin of error (10%) of heat capacity measurements.

Figure 7c shows the thermal conductivity of both samples as a function of temperature, calculated with eq. 1. In both cases, the thermal conductivity is decreasing with increasing temperature, because of Umklapp scattering [44]. At room temperature, the thermal conductivity of the 2 mol% K-excess ceramics is around 10 W m$^{-1}$ K$^{-1}$, like values reported in the literature for single crystals, which range between 10 and 12 W m$^{-1}$ K$^{-1}$ (refs. 10,11,45). The value for the ceramics without K-excess is lower, around 8 W m$^{-1}$ K$^{-1}$. Since both ceramics have similar densities and final phases compositions, this difference is attributed to the scattering by grain boundaries which is stronger for the stoichiometric ceramics where the grain size is below 1 µm, while grains expand up to 5 µm in the 2 mol% excess K ceramics. Similar phenomena have been reported before in La-doped SrTiO$_3$ where at room temperature the thermal conductivity decreases from 9 W m$^{-1}$ K$^{-1}$ for 6 µm grain sizes to 5 W m$^{-1}$ K$^{-1}$ for 0.1 µm grain sizes [46]. In KTaO$_3$, at higher temperatures, both thermal conductivities converge towards 4 W m$^{-1}$ K$^{-1}$. The overall trend is consistent with measurements reported on single crystals, up to 570 K [43].

It is interesting to compare these results with thermal conductivity values reported for a system close in composition, such as KNbO$_3$. Experimentally, the thermal conductivity of KTaO$_3$ single crystals is decreasing with increasing Nb content up to 37% [43]. Here, we find that the thermal conductivity of KTaO$_3$ at room temperature is around 8-11 W m$^{-1}$ K$^{-1}$, instead of 2.5 W m$^{-1}$ K$^{-1}$ for KNbO$_3$ (Fig. 7c). Considering only the mass of the elements, since tantalum is heavier than niobium, the thermal conductivity of KTaO$_3$ should increase when Ta is replaced by Nb [47]. Instead, the opposite trend is observed. Several phenomena can combine to account for this behaviour. First, KTaO$_3$ stays cubic while KNbO$_3$ undergoes a series of transitions in the temperature range investigated (orthorhombic, tetragonal) and is cubic only above 700 K [48]. The thermal conductivity of orthorhombic and tetragonal phases is expected to be lower than the thermal conductivity of cubic phases [49]. In addition, KNbO$_3$ is ferroelectric, while KTaO$_3$ is not. The presence of ferroelectric domains may reduce further the thermal conductivity [50]. First-principle based theoretical calculations also reveal that the Grüneisen parameter is larger in KNbO$_3$ compared to KTaO$_3$, indicating a strong phonon anharmonicity in KNbO$_3$, which results in a reduced thermal conductivity [51].

## 4. Conclusion

We have shown that spark plasma sintering is a suitable technique to obtain dense and single-phase $KTaO_3$ ceramics through a fast synthesis process. The addition of 2 mol% K-excess in the initial powder helps lowering sintering temperatures and pressures. The best SPS parameters with K-excess are found to be 60 MPa of pressure between the punches, for 30 seconds at 1243 K. The best parameters for powders without K-excess are found to be 100 MPa, for 1 minute at 1423 K, although polishing is necessary to remove the superficial $K_6Ta_{10.8}O_{30}$ secondary-phase. Oxidation of the samples after SPS to remove oxygen vacancies is performed at 1273 K for 1 hour. The initial K-excess powder leads to higher relative densities and larger grains in the spark plasma sintered ceramics.

Dielectric measurements reveal different behaviours for the two samples, with a maximum at low temperature of 4000 for the sample with an initial excess of 2 mol% in potassium, comparable to the value of high-quality single crystal $KTaO_3$ and correlated to the microstructure, especially the large grain size and reduced amount of grain boundaries in this K-excess composition. Thermal conductivities are found to be much higher than in $KNbO_3$.

## 5. Declaration of Competing Interest

The authors declare that they have no known competing financial interests or personal relationships that could have appeared to influence the work reported in this paper.

## 6. Acknowledgments

The authors are grateful to Tatiana Chartier and Jacob Baas for technical support, Beatriz Noheda for giving access to the low-temperature dielectric spectroscopy setup, and Kevin Nadaud for discussions on dielectric spectroscopy measurements. EKHS is grateful to EPSRC for support (grant EP/P024904/1). Gaurav Vats acknowledges funding from the European Union's Horizon 2020 research and innovation programme under the Marie Skłodowska-Curie Grant Agreement No. 892669. Co-funded by the European Union (ERC, DYNAMHEAT, N°101077402). Views and opinions expressed are however those of the authors only and do not necessarily reflect those of the European Union or the European Research Council. Neither the European Union nor the granting authority can be held responsible for them.

| | Temperature (K) | Pressure (MPa) | Heating rate (K min$^{-1}$) | Dwell time (min) | $d_{geo}$ (g cm$^{-3}$) | Relative $d_{geo}$ | $d_{Archimedes}$ (g cm$^{-3}$) | Relative $d_{Archimedes}$ | XRD phases after SPS |
|---|---|---|---|---|---|---|---|---|---|
| $K_{1.02}TaO_3$ | 1223 | 60 | 100 | 2 | 6.57 | 94% | 6.60 | 94% | $KTaO_3$ |
| | **1243** | **60** | **100** | **0.5** | **6.83** | **97%** | **6.80** | **97%** | **$KTaO_3$** |
| $KTaO_3$ | 1273 | 60 | 100 | 1 | 5.43 | 77% | 5.46 | 78% | $KTaO_3$ |
| | 1373 | 60 | 100 | 1 | 5.57 | 79% | 5.69 | 81% | $KTaO_3$ |
| | 1373 | 100 | 100 | 1 | 6.05 | 86% | 6.09 | 87% | $KTaO_3$ |
| | **1423** | **100** | **100** | **1** | **6.35** | **90%** | **6.67** | **95%** | **$KTaO_3$ +$K_6Ta_{10.8}O_{30}$ surface** |
| | 1473 | 100 | 100 | 1 | 6.82 | 97% | 6.85 | 98% | $KTaO_3$ +$K_6Ta_{10.8}O_{30}$ bulk |

*Table 1: Investigation of SPS parameters for 0 mol% K-excess and 2 mol% K-excess KTaO$_3$ powders. Lines in bold indicate ceramics used for X-rays, SEM, dielectric and thermal measurements.*

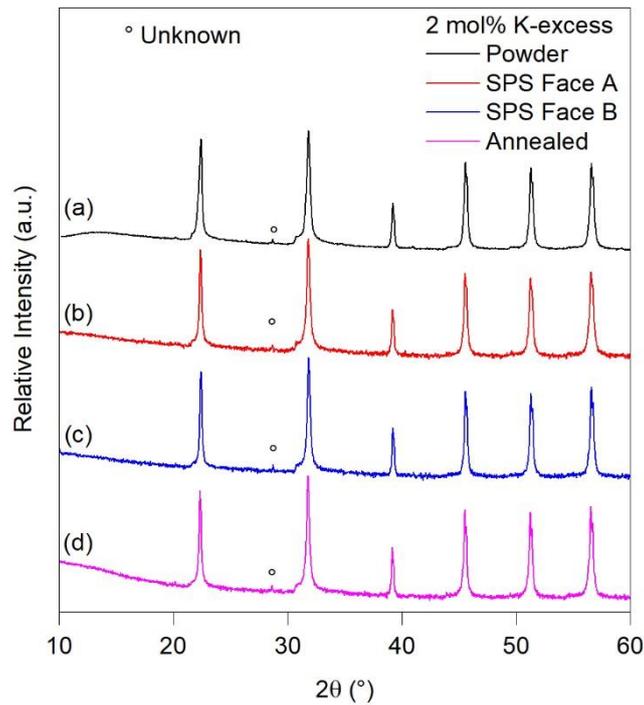

*Fig. 1: X-ray diffraction patterns (logarithmic scale) of the (a) 2 mol% K-excess KTaO$_3$ powder, faces (b) A and (c) B of the ceramics after spark plasma sintering at 1243 K and (d) after annealing under oxygen flow. The peak at 28.5° remains in all these diffractograms and is not attributed to KTaO$_3$. It also appears in (K,Na)NbO$_3$ phase [52].*

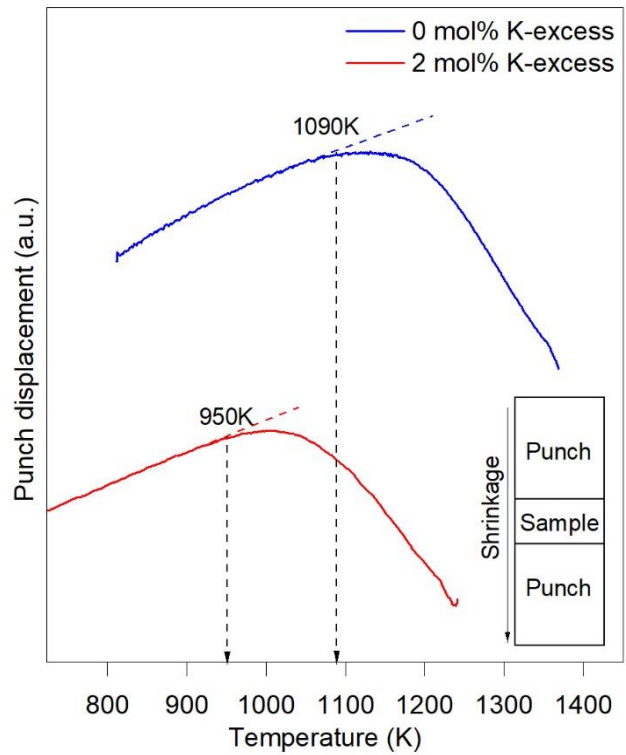

*Fig. 2 Shrinkage curves of 0 mol% K-excess ($T_{sintering}$=1243 K) and 2 mol% K-excess ($T_{sintering}$=1423 K) $KTaO_3$ powders during the spark plasma sintering process. Y-axis represents the punch displacement.*

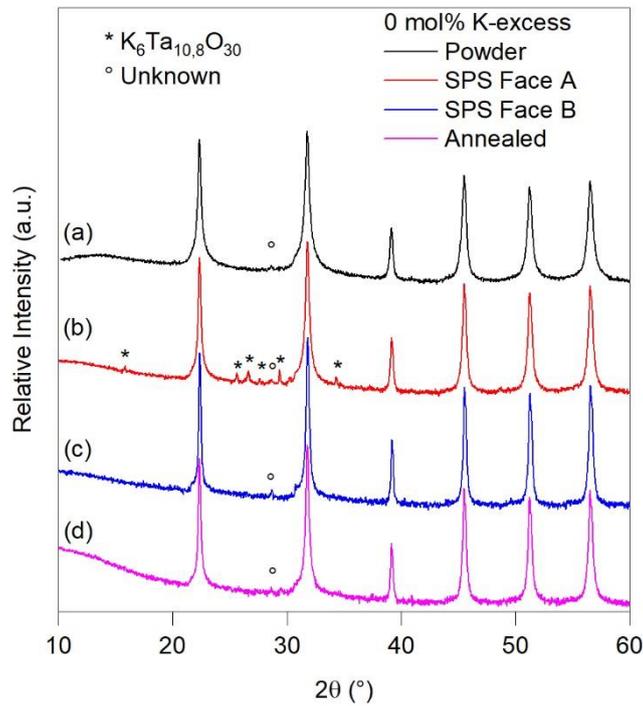

*Fig. 3 X-Ray Diffraction (logarithmic scale) of (a) 0 mol% K-excess $KTaO_3$ powder, faces (b) A and (c) B of the ceramics after spark plasma sintering at 1423 K and (d) after annealing under oxygen flow. The peak at 28.5° remains in all these diffractograms and is not attributed to $KTaO_3$. It also appears in $(K,Na)NbO_3$ phase [52].*

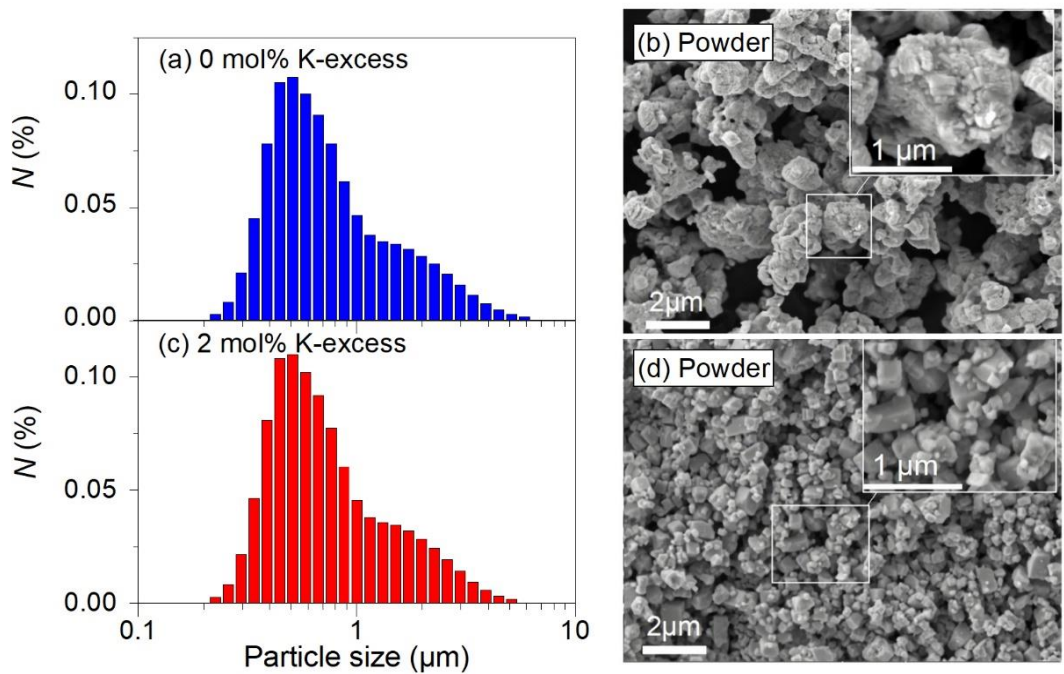

*Fig. 4 (a,c) Studies of the particle size in both powders by laser granulometry with N the distribution of particle size in percentage. SEM images of the powders with (b) 0 mol% K-excess and (d) 2 mol% K-excess $KTaO_3$.*

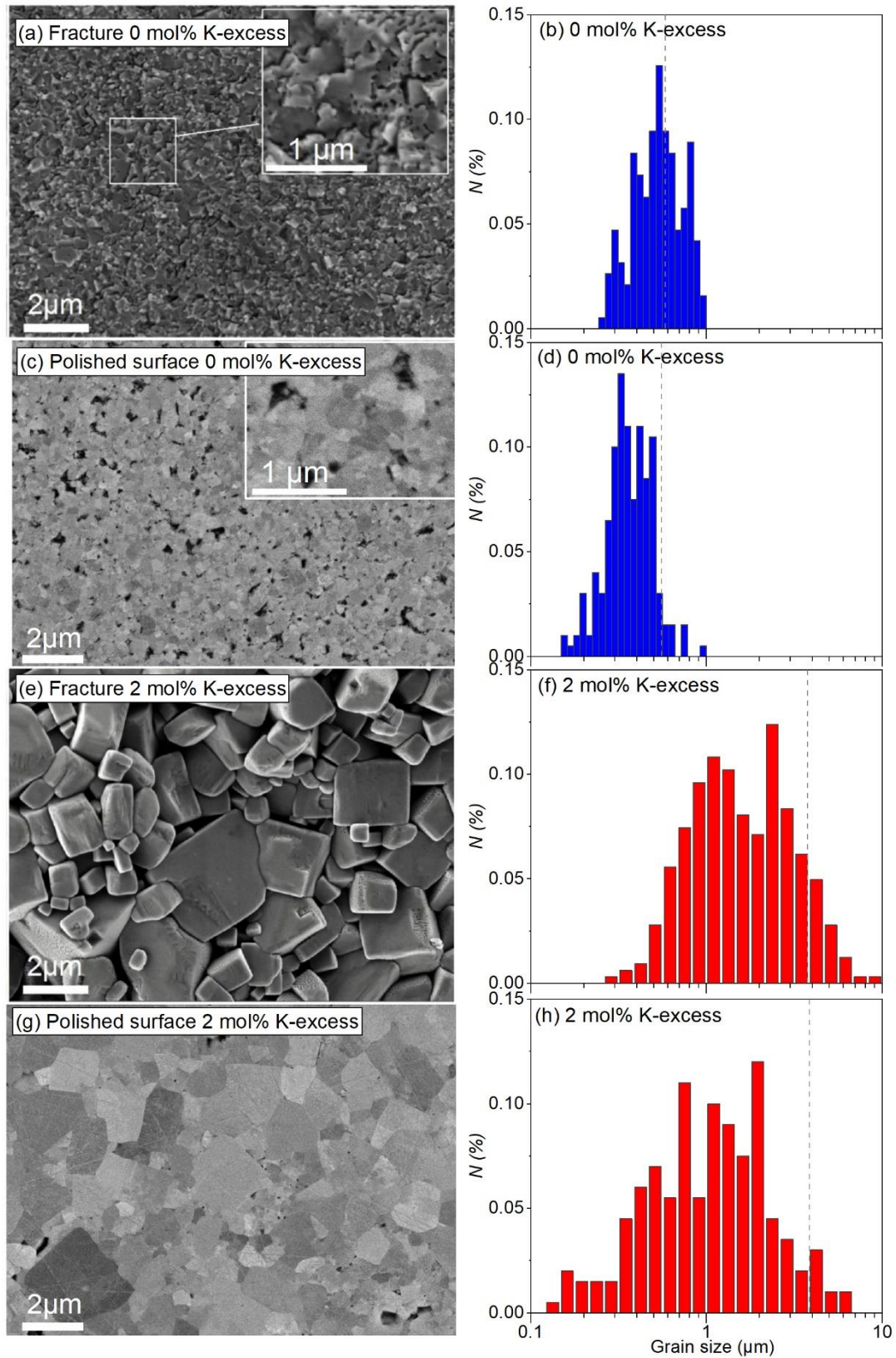

*Fig. 5 SEM images of fractured $KTaO_3$ ceramics with (a) 0 mol% K-excess and (e) 2 mol% K-excess, with (b,f) grain size distributions for both images. SEM images of the polished surfaces of the same ceramics with (c) 0 mol% K-excess and (g) 2 mol% K-excess, with (d,h) grain size distributions for both images. Vertical dashed lines in (b,d,f,h) indicate the mean grain size measured with the linear intercept method. In (d) and (h), all values have been multiplied by a proportionality constant of 2.25 to give an estimate of the real grain size [53].*

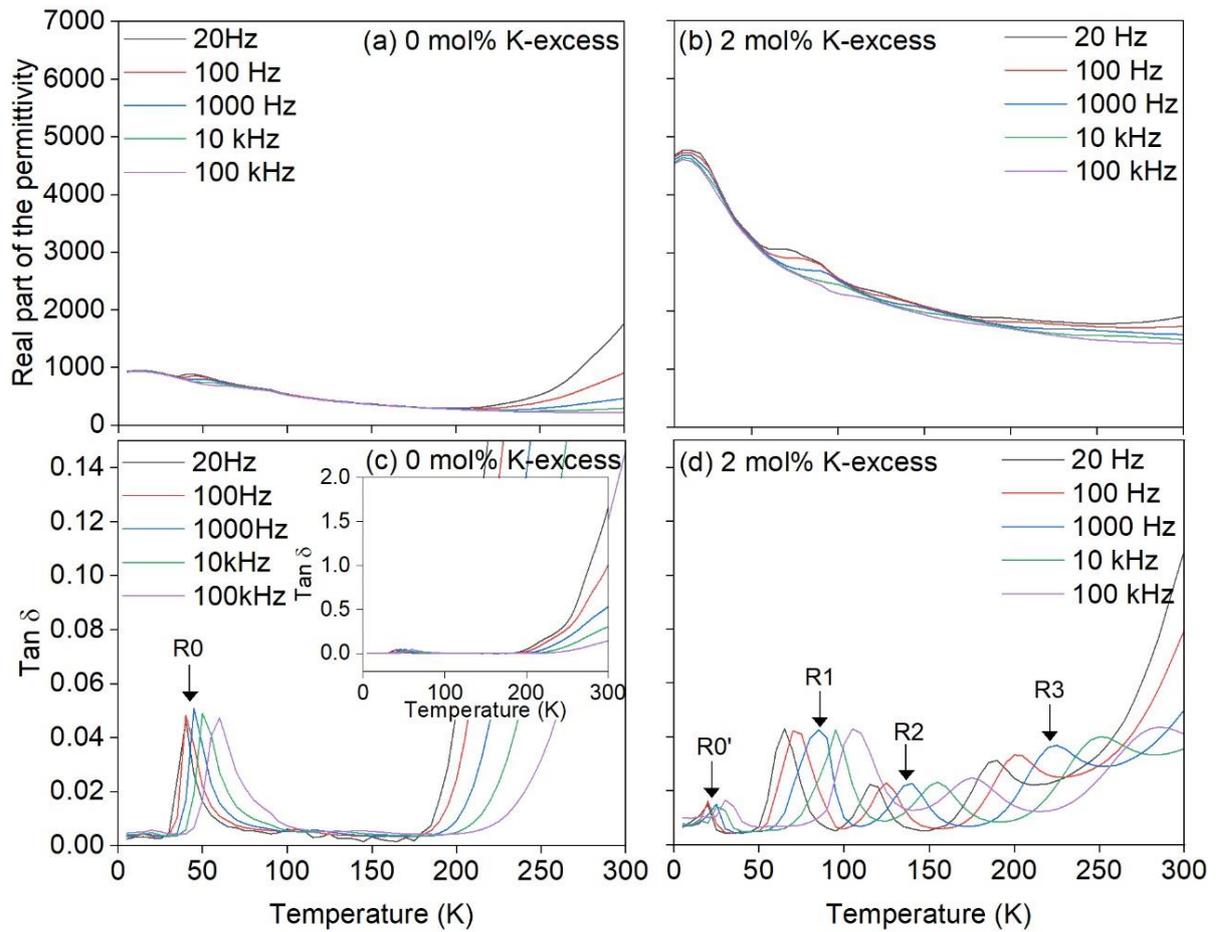

*Fig. 6 Dielectric measurements at low temperatures (5K-300K) with the real part of the dielectric permittivity of (a) 0 mol% K-excess and (b) 2 mol% K-excess $KTaO_3$ and (c,d) dielectric losses of both samples at different frequencies.*

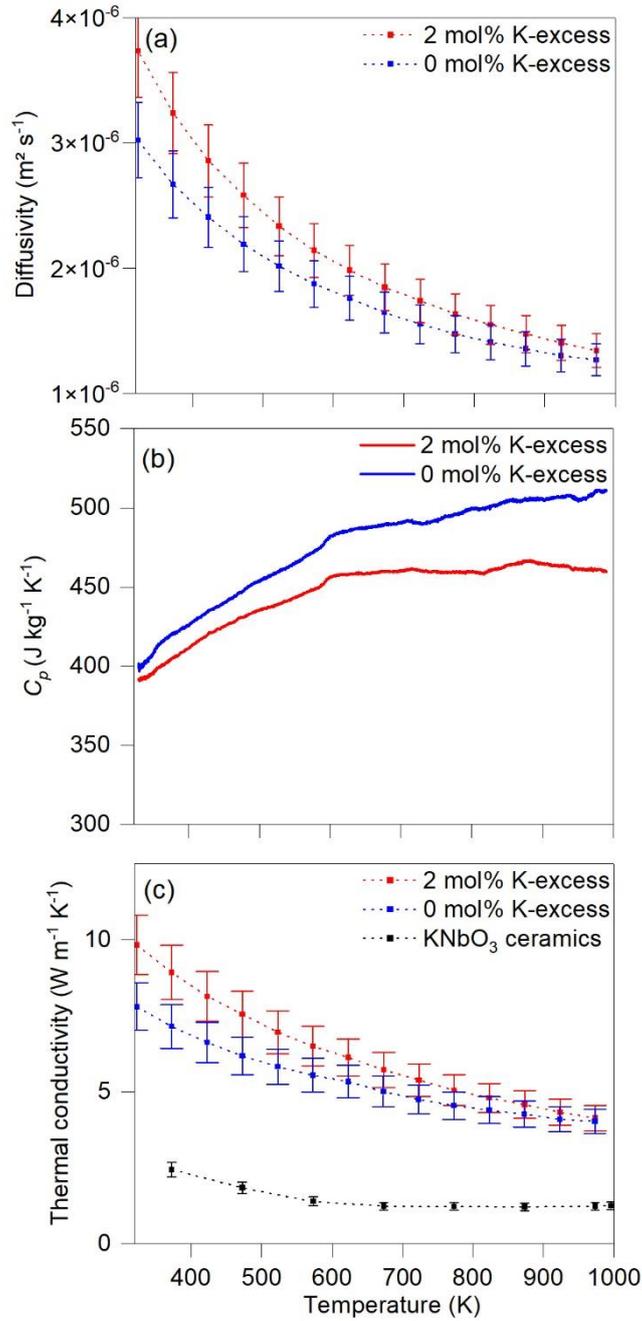

*Fig. 7 (a) Thermal diffusivity and (b) specific heat capacity of dense $KTaO_3$ ceramics, with 0 mol% K-excess and 2 mol% K-excess. (c) Thermal conductivity of dense $KTaO_3$ ceramics, with 0 mol% K-excess and 2 mol% K-excess (this work), and $KNbO_3$ ceramics (96% relative density, 1-4μm grain size) [9] as a function of temperature. Dashed lines are guides to the eyes.*